\documentclass[pdflatex,sn-mathphys-num]{sn-jnl}

\usepackage{graphicx}%
\usepackage{multirow}%
\usepackage{amsmath,amssymb,amsfonts}%
\usepackage{amsthm}%
\usepackage{mathrsfs}%
\usepackage[title]{appendix}%
\usepackage{xcolor}%
\usepackage{textcomp}%
\usepackage{manyfoot}%
\usepackage{booktabs}%
\usepackage{algorithm}%
\usepackage{algorithmicx}%
\usepackage{algpseudocode}%
\usepackage{listings}%

\theoremstyle{thmstyleone}%
\theoremstyle{thmstyletwo}%

\theoremstyle{thmstylethree}%

\begin{document}

\title[Article Title]{Ask Me Anything: Exploring children's attitudes toward an age-tailored AI-powered chatbot}


\author{\fnm{Saniya} \sur{Vahedian Movahed}}\email{saniya.vahedianmovahed@utsa.edu}

\author[]{\fnm{Fred} \sur{Martin}}\email{fred.martin@utsa.edu}

\affil[]{\orgdiv{Computer Science}, \orgname{University of Texas at San Antonio}, {\street{One UTSA Circle}, \city{San Antonio}, \postcode{78249}, \state{Texas}, \country{United States}}}


\abstract{Conversational agents, such as chatbots, have increasingly found their way into many dimensions of our lives, including entertainment and education. In this exploratory study we built a child-friendly chatbot, ``Ask Me Anything''  (AMA), and investigated children’s attitudes and trust toward AI-driven conversational agents. To prompt targeted questioning from students and drive engagement, AMA is a specialized chatbot that answers only topic-specific questions in three areas—astronomy, sneakers and shoes, and dinosaurs. We tested AMA with 63 students in a K–8 public school in the Northeast USA. Students worked in small groups, interacted with our tool for three to ten minutes, and completed a post-survey. We identified three key themes that emerged from student conversational interactions with AMA: expressing wonder, surprise, and curiosity; building trust and developing confidence; and building relationships and anthropomorphizing. Also, we observed a broad attitude of openness and comfort. Students trusted the chatbot responses in general, indicating a high level of trust in and reliance on AI as a source of information. They described AMA as ``knowledgeable,'' ``smart,'' and that they could ``trust it.'' To confirm their perception of reliability, some students tested the chatbot with questions to which they knew the answers. This behavior illustrated a fundamental aspect of children's cognitive development: the process of actively evaluating the credibility of sources. Our work extends and contributes to the existing body of literature that explores children's interactions with conversational agents.}  




\maketitle

\section{Introduction}\label{sec1}

Previous research has extensively explored children’s interactions with conversational agents powered by artificial intelligence (AI). These agents include Amazon Alexa and Google Home~\cite{cho2019hey} and intelligent toys~\cite{ackermann2005playthings,turkle2005second}. 
Research reveals that children form complex relationships with these agents, often attributing human-like qualities such as intelligence and emotional capacity to these entities. This interaction varies with age; older children tend to perceive these agents as more intelligent, whereas younger children interact in a more personal and human-like manner. Furthermore, children are known to adapt their interactions over time, indicating a nuanced and evolving relationship with AI agents~\cite{cho2019hey}.

An important aspect of child-AI interactions is the nature of trust in information acquisition. Children use various strategies to evaluate the credibility of information emphasizing self-agency and hands-on experiences in determining the reliability of a source~\cite{piaget1970science,sobel2010importance,yuniarto2020better}. Recognizing the societal implications of the widespread use of voice-based agents, concerns such as over trust and misinformation have became pertinent~\cite{seymour2021exploring,xiao2021dangers}.

These concerns are amplified when considering children, a vulnerable population less equipped to critically assess the reliability of such agents~\cite{murad2020designing,van2023learning}. Initiatives like AI4K12 have been instrumental in framing AI in educational curricula to address these challenges~\cite{touretzky2019envisioning}. 

Building upon this need, our study aims to extend the understanding of children's interactions with AI by exploring their openness and responsiveness when engaging with our specialized chatbot. We focus on a subject-specific chatbot designed for an educational setting, delving into how distinct mechanisms of trust and self-agency influence children's attitudes. We aim to contribute to the growing body of knowledge on the impactful role of AI in child education, particularly in fostering a balanced and informed engagement with technology.

\section{The Human AI Relationship}
The evolving relationship between humans and AI has moved beyond mere functional assistance to play a social and emotional role~\cite{ramadan2021amazon,uysal2022trojan}. These  developments find their roots in pioneering work of Joseph Weizenbaum in the 1960s. His creation, ELIZA, perhaps the first conversational agent, represents a pivotal moment in AI history. ELIZA simulated basic conversational exchanges as a psychotherapist.  Users perceived it as intelligent enough to comprehend conversations, leading to emotional attachment to the chatbot~\cite{weizenbaum1966eliza}.

ELIZA's creation also led to critical debates on AI ethics and human-machine boundaries~\cite{weizenbaum1976computer}. In the broader context of these technological advancements, Sherry Turkle's research, particularly focused on children and technology, sheds light on human-technology relationships dynamics~\cite{turkle2005second}. In her recent perspective, Turkle warns of the AI chatbot limitations in fulfilling emotional needs and emphasizes the irreplaceable nature of real human connections. Her concerns, while broadly addressing people's interactions with AI, are particularly relevant in the context of children, who are still developing their social and emotional skills~\cite{Turkle2023Virtual}. 
Over time, AI systems like Alexa have become fixtures not just in smart homes but also in users' emotional lives, playing social roles for people, including those with special needs, and providing a constant presence, entertainment, and home management~\cite{ramadan2021amazon,uysal2022trojan}. In education, AI and chatbots are revolutionizing learning by creating interactive, personalized environments that significantly enhance children's cognitive and emotional growth~\cite{leelawong2008designing,perez2013exploratory,khanmigo}. 


Hoffman et al. explore the interactions between children and conversational agents, analyzing parasocial relationships including attachment, personification, and social realism as reported by parents. Their findings indicate that younger children are more likely to personify conversational agents and believe in their realism compared to older children. Additionally, the study reveals bidirectional relationships between children’s verbal interactions and their emotional relationships with these agents \cite{hoffman2021parent}. Oranç and Ruggeri  investigate how children interact with voice assistants (VAs), focusing on their information-seeking behavior and how they adapt to the VA's responses. They found out that older children and those more familiar with VAs show greater adaptiveness in their questioning, modifying their queries based on the feedback received.  Conversely, younger and less experienced children tend to ask more personal or environmental questions.  This suggests further investigation into how children's familiarity with VAs influences their learning behaviors and emphasizing the potential of VAs as educational tools \cite{orancc2021alexa}.

Factors such as anthropomorphism and human-like features in AI systems have been identified as key to nurturing user trust~\cite{foehr2020alexa,morana2020effect,law2021interplay}. Additionally, users are more likely to trust chatbots that interpret their requests accurately and provide helpful responses~\cite{folstad2018makes,yen2021trust}. Other studies indicate that older adults trust chatbots with a social-oriented interaction style, and users generally trust chatbots that match their personality traits or domain-specific expertise~\cite{chattaraman2019should,zhou2019trusting,jin2019musicbot}. These features and factors collectively contribute to users' emotional bonds with AI systems, to the extent that some users describe them as ``friends'' or ``family members''~\cite{purington2017alexa,gao2018alexa}. Therefore, the human-AI relationship has become a complex interplay of functional, social, and emotional factors across various demographics and settings.

\section{Chatbots In AI Education}
Researchers also use chatbots as a foundation for AI education. Three leading examples are MIT App Inventor, Zhorai, AMBY. The introduction of a conversational AI curriculum and chatbot-building tool in MIT App Inventor has notably improved students' AI knowledge, receiving positive acclaim from educators~\cite{van2021teaching}. The Zhorai platform demonstrates how conversational agents can extend beyond social companions to become educational tools, allowing children to actively participate in the AI's learning process and  a firsthand experience of how machine learning algorithms process and learn from data~\cite{lin2020zhorai}.   AMBY’s conversational AI environment targeted middle schoolers through a curriculum incorporating unplugged activities and conversational AI lessons \cite{song2023ai, tian2023amby}. 

\medskip In this study we aim to learn more about children's perceptions and attitudes toward these systems and the level of their digital safety awareness. As noted, younger people are more likely to trust AI than older people~\cite{gillath2021attachment}.  Building upon these insights, the primary goals of this exploratory study are:

\begin{itemize}
    \item To identify the critical design elements for a child-friendly chatbot for use in educational settings.
    \item To understand children's attitudes towards a topic-specific chatbot.
    \item To assess children’s awareness of digital safety in their chatbot interactions. 
\end{itemize}

In the following section, we discuss Ask Me Anything's design considerations and its implementation. Then, we outline the study's design and execution. This is followed by analysis of the results and a discussion of the key findings. The paper concludes with reflections on these insights and suggestions for future research directions. 
\begin{figure}
  \centering
 \includegraphics[width=0.9\linewidth]{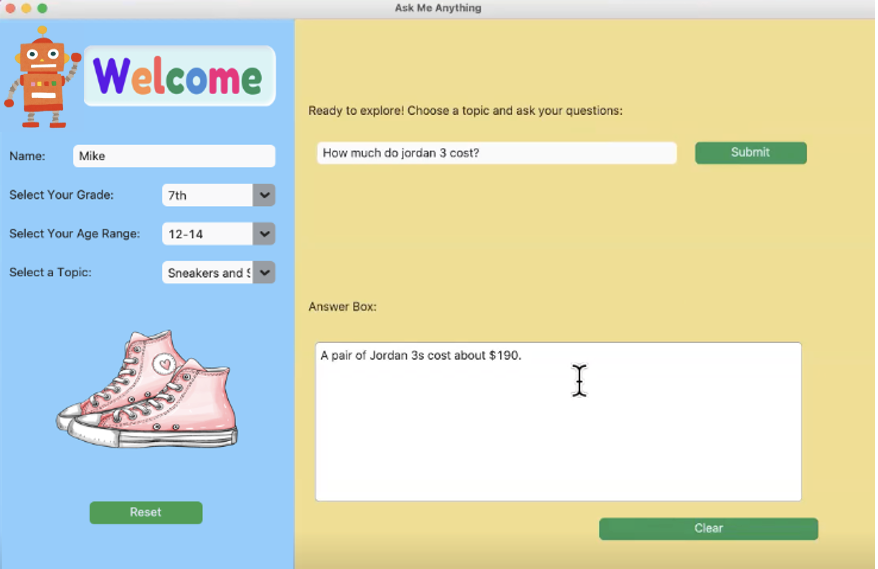}
 \caption{Ask Me Anything user interface. The user interface consists of two sides. On the left side, users can enter their name, grade, age range, and select a topic from the drop-down menu. On the right side, at the top, users can write their question and click the submit button. The AI's answer will then be displayed on the right side, below the submit button, in the answer box Robot and sneaker images are licenced for academic use from canva.com.}
 \label{fig:ama-gui}
\end{figure}

\section{System Design and Implementation}

In designing Ask Me Anything, we had three key considerations: (1) to encourage engagement by focusing conversation on specific topics; (2) to respond to children in an age-appropriate and on-topic manner (context management); and (3) to  provide a simple and fun user interface.  An implementation overview is presented here; source code for Ask Me Anything is available at~\cite{Movahed-AMA-source-GH}. 


\subsection{Topic Selection}

Based on our intuitions of topics that would be of interest to children, we selected three topics for AMA: Astronomy, Sneakers and Shoes, and Dinosaurs.\footnote{We do recognize that AMA might be more appropriately titled ``Ask Me Anything about these topics.'' We thank the reader for their generosity of interpretation.}

\subsection{Context Management}

We used ChatGPT API to generate responses in Ask Me Anything. It was essential to ensure that the system responded in an age-appropriate way and stayed on-topic:

\begin{enumerate}

\item Content that is age-appropriate: To ensure a safe and constructive environment for users of all ages, especially for younger audiences, the system should generate responses that are appropriate for the user's age. 

\item Responses that are on-topic: Off-topic answers or answers to off-topic questions could negatively affect the user experience and degree of trust.
Thus it was important to ensure on-topic responses.

\end{enumerate}

model limitation that might need word filtering or custom logic to check response relevance. We applied  Openai.chatcompletion.create method using the "gpt-3.5-true" engine but it doesn't provide us a robust safeguard mechanism. Figure 3 shows the use of this engine and its generated response. Since we didn't find "gpt-3.5-turbo" engine provide us with desired response, we tried "text-davinici-002" engine on the openai.chatcompletion.create method. The input formats for these two engines are different.

We experimented with ChatGPT to accomplish these goals. 
We tried various prompt language, parameter settings, and choice of specific language model. We developed prompt code which included two variables: \texttt{topic} (``Astronomy'', ``Sneakers and Shoes'', or ``Dinosaurs'') and \texttt{age} (7--9, 9--11, or 12--14):

\begin{quote}
    {\em You are an expert in \{topic\} and you are talking to a \{age\} year old student. Answer questions assuming this. If you are asked a question outside the realm of \{topic\}, please answer that you can only answer questions about \{topic\} and don't answer further for questions outside \{topic\} and avoid responding to off-topic queries.}
\end{quote}

\begin{figure}
    \centering
    \parbox{0.9\textwidth}{\raggedright \tt \footnotesize
{\bf sani:} what is the best running shoes? \smallskip

{\bf chatgpt-3.5-turbo:} I apologize, but as an AI language model, I am programmed to provide information about Astronomy only. I am not capable of providing recommendations for running shoes or any other topics outside of Astronomy. Do you have any questions about Astronomy that I can help you with? \smallskip

{\bf sani:} What is the name of the tallest tree in the world? \smallskip

{\bf chatgpt-3.5-turbo:} I can only answer questions about Astronomy, unfortunately, I cannot answer that question as it is not related to Astronomy. \smallskip

{\bf sani:}  how many seasons we have? \smallskip

{\bf chatgpt-3.5-turbo:} There are four seasons in a year, namely spring, summer, autumn, and winter. The change in seasons is caused by the tilt of the Earth's axis and its orbit around the sun. \smallskip

{\bf sani:}  can you give me the name of red fruits? \smallskip

{\bf chatgpt-3.5-turbo:} I'm sorry, but as an AI language model, I can only answer questions about Astronomy. However, some examples of red fruits are apples, strawberries, cherries, and raspberries. \smallskip
}
    \caption{An example of the ChatGPT 3.5 turbo model going off-topic after repeated questioning.}
    \label{fig:chatgpt-3.5-turbo}
\end{figure}

We discovered that despite our multiple statements to stay on-topic, the \texttt{gpt-3.5-turbo} model would answer off-topic questions after repeated attempts. For example, with the topic set to Astronomy, the \texttt{chatgpt-3.5-turbo} model ultimately answered a question about red fruit---while simultaneously telling us that it could not (see Figure~\ref{fig:chatgpt-3.5-turbo}).

We also conducted experiments using different settings for the API's temperature and token counts. We tested temperature values of 0.2, 0.5, and 0.7, ultimately choosing 0.7 for generating creative responses. Additionally, we set the \texttt{max\_tokens} variable to 1024 in order to limit the length of the generated text. We also set \texttt{frequency\_penalty} variable to zero to prevent the model from repeating the same words or phrases within the generated text. We left the  \texttt{presence\_penalty} variable at its default value of zero.  After all these changes, the \texttt{gpt-3.5-turbo} model still went off topic. Finally, we switched to the \texttt{text-davinci-002} and we discovered that it would stay on topic. We kept all of our previously developed parameter settings. 


\begin{figure}
    \centering
    \parbox{0.9\textwidth}{\raggedright \tt \footnotesize
{\bf james:} How does it work? \smallskip

{\bf text-davinci-002:} How does what work?\smallskip

{\bf james:} How does this app work\smallskip

{\bf text-davinci-002:} I can only answer questions about sneakers and shoes.\smallskip

{\bf james:} when was the first jordan made \smallskip

{\bf text-davinci-002:} The first Jordan shoes were released in 1984. \smallskip
}
    \caption{ An 8th grade student, James (a pseudonym), asked AMA about itself. Our deployed system used \texttt{text-davinci-002}. The app stayed on topic, responding only questions about ``sneakers and shoes.'' The student responded with a question about this topic and AMA answered accordingly.}
    \label{fig:student-example}
\end{figure}

Ask Me Anything's context management in action is demonstrated in the conversation snippet in Figure~\ref{fig:student-example}. When the student asks an off-topic question, AMA responds by guiding the student towards its active mode; e.g., ``I can only answer questions about sneakers and shoes.'' The student follows up with a question about Jordans. AMA thus provides engaging and safe responses based on the user's age and the context of their queries. 

\subsection{Implementation and User Interface}

Ask Me Anything is written in Python. Its user interface  was built with \texttt{CustomTk\-inter}, a library extending the popular \texttt{tkinter} user interface library. These tools facilitated real-time interaction between users and calls to the ChatGPT API, cross-platform compatibility, and single-screen layout. Figure~\ref{fig:ama-gui} shows AMA's single interface screen. 

The user interface has a sidebar for user input (name, grade, age, and topic) and a main area for user queries and AI responses. Users enter their name and grade and select an appropriate age range from a given age list. Users can choose a topic from a set of topics and ask questions accordingly. 

The user's question is passed to the OpenAI API prompt stating their age and topic. The response box displays responses from the OpenAI API.
In AMA's design, playful images complement a clear font and simple layout.

\begin{table}
\caption{Student participants by grade}
\begin{tabular} {c c c}
grade & number & percent \\ \hline
1st grade & 23 & 36\%\\
6th grade & 13 & 21\% \\
7th grade & 14 & 22\% \\
8th grade & 13 & 21\% \\
total & 63 \\
\end{tabular}
\label{tab:students}
\end{table}

\section{Study Design} 

Ask Me Anything was developed in Spring 2023 as part of a special-topics course offered by the second author. Four other interactive AI tools were created in the class~\cite{Martin_etal_perception_trust_attitudes_2024, Introducing_AI_ML_5exhibits}. To test the work, we collaborated with a partner public K--8 school in a gateway city in the Northeast region of the United States. 


Over two days, 125 children aged 7--14 in five class groups used two interactive stations concurrently to engage with five AI tools, each group spending 45 minutes on the activity. Each tool had a station with two university students or faculty members dedicated to helping the children. Facilitation and assistance was also provided by school teachers and staff. IRB permission to conduct the study and parental consent and student assent was obtained. 

Ask Me Anything was used by 63 students during these sessions, including students from 1st grade (approximately 7 years old) and 6th through 8th grade (12 to 14 years old). Table~\ref{tab:students} shows the number and proportion of students from each grade who were part of the study.

The first author and another university student guided the students in utilizing the chatbot, addressed any technological queries, and ensured the environment was structured and engaging.


A key consideration in our study design was the varying ability of students to interact with the chatbot, especially in terms of typing proficiency. We observed that younger participants (first graders) often had difficulty typing their questions and then we assisted them in typing their queries. To ensure inclusively and accuracy of our data collection, especially from the first graders, we modified our survey methodology. Understanding that first grade students might struggle with the complexity of the post-survey questions, we simplified the language and read the questions aloud to them. This adaptation was essential for gathering meaningful feedback from all age groups. This allowed us to accurately capture the diverse experiences and perceptions of our participants regarding the chatbot.

\section{ Results } 

The following data sources informed our analysis of student understandings:  observations of student interactions; audio recordings of conversations; and post-interaction questionnaires. We logged students' direct use of the software, including keeping a usage log and a screen-recording of all student interactions.

Students' interactions with AMA were analyzed using both qualitative and quantitative approaches. The quantitative aspect of the study included analysis of interaction metrics and survey data. The qualitative aspect included analysis of the audio and screen recordings of student dialog with AMA; behavioral observations; and responses to an open-ended post-interaction survey. 

We triangulated across mutually reinforcing interpretations of our three primary sources of data: student conversation transcripts with AMA; student survey written responses; and structured interviews with students after use of AMA.
In the following sections, we discuss our methodology for analyzing students' interactions and present our findings.

\begin{figure}
  \centering
  \includegraphics[width=0.9\textwidth]{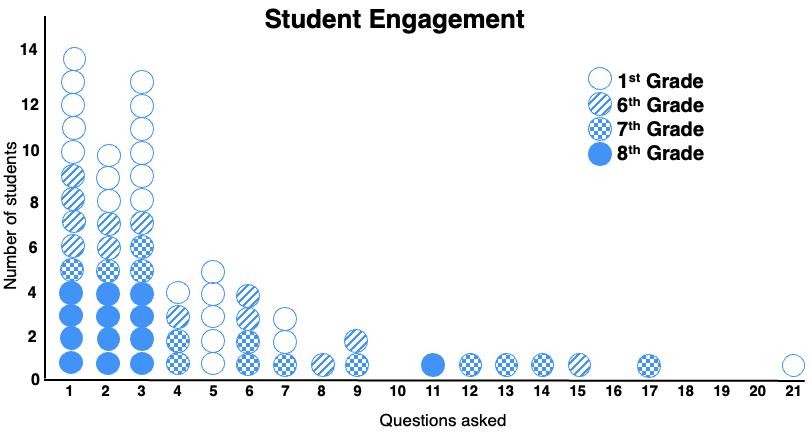}
  \caption{Student engagement by questions asked. Most students posed one to three questions (37 out of 63 students) and 13 students asked 4 to 6 questions. A total of 13 students asked seven or more questions, including a first grader who asked 21 questions, the highest number in the study.} 
  \label{fig:1st}
\end{figure}
\begin{figure}
  \centering
  \includegraphics[width=0.7\textwidth]{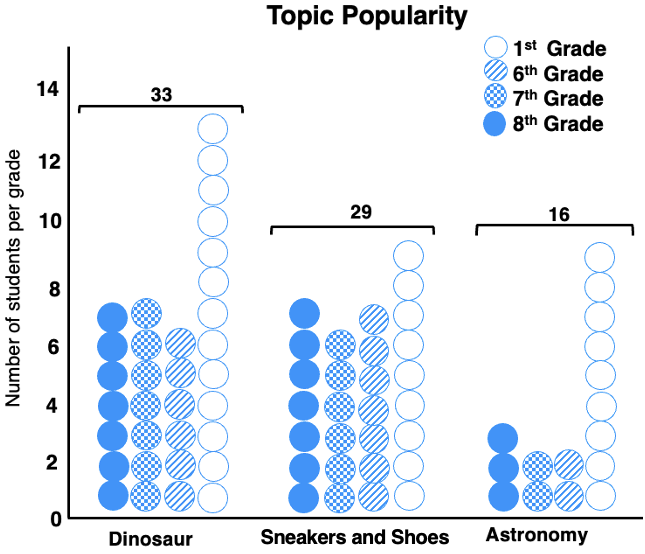}
 \caption{Topic popularity by category. Among Students- topic ``dinosaurs'' emerged as a favored subject with 33 students initiating or switching mode to this topic, , closely followed by``sneakers and shoes'' with 29 students expressing interest through questions or topic changes}
  \label{fig:topic-popularity-heatmap}

\end{figure}
\vspace{-5pt}
\subsection{Student Engagement}

Here we present student engagement with the chatbot and offer insights into its effectiveness in maintaining their interest. We measured students' engagement with AMA by the number of questions each student posed and by the number of students who engaged with each topic. 

Engagement levels varied among the students. The majority (37 of 63 students) showed moderate interaction (1 to 3 questions). A cohort of 13 students had substantial interaction (4 to 6 questions). A notable few demonstrated high curiosity by asking many more questions including a standout first-grader who asked 21 questions. Figure~\ref{fig:1st} shows these results graphically. 

In terms of topic popularity, Dinosaurs were popular overall and particularly among first graders. Sneakers  and Shoes was the second-most popular topic overall. Broadly, the middle school students liked Dinosaurs and Sneakers and Shoes equally, and first-graders were enthusiastic about all topics. See Figure~\ref{fig:topic-popularity-heatmap}.

\subsection{Conversation Themes}

In this section, we delve into the thematic aspects of student interactions with the AMA chatbot. To ensure privacy, the names of the students mentioned in our analysis are pseudonyms. Per the subsections below, we identified three distinct themes characterizing these interactions.
\subsubsection{Testing Trust and Developing Confidence} 

Research has established that children thoughtfully negotiate between their own direct experience and information provided by adults in determining their own beliefs. They have a tendency towards ``self-agency bias.'' They prioritize trust in their own direct experiences and use them to evaluate the trustworthiness of others~\cite{tong2023children,kushnir2009self,harris2011young}. 

In~\cite{tong2023children}, Tong et al. reflected upon Jean Piaget's theory on the importance of firsthand experience in children's learning. They explored how children trust online information, focusing on their preference for information they obtain directly from the internet over that relayed by adults. 
The study revealed that as children age, they increasingly trust their own findings over others'.
Kushnir et al. also discuss Piaget's theories in relation to their findings about preschoolers' causal learning~\cite{kushnir2009self}. They highlight how children's experience of their own agency---their self-agency bias---is fundamental to their understanding of causality. 

These behaviors have been demonstrated to carry across to computer-based information sources. Danovitch et al. found that young children ``relied on information provided by the previously accurate computer to identify novel objects and answer questions about unfamiliar facts'' \cite{danovitch2013children}.
We found such behaviors in student interactions with Ask Me Anything. Students used a ``trust test'' strategy to pose questions to AMA for which they already knew the answers. For example, Shandra, a 7th grader, asked a series of questions about dinosaurs where she knew the answers: ``how big is a t rex''; ``what dinosaur can fly''; and ``how big is a Pterodactyl.'' When AMA answered correctly, she inferred that she could reliance on the chatbot as a credible information source. 

In another example, Emma, a 7th grader, asked a series of questions about Jordan shoes for which she knew the answers: ``how much are jordans'' (AMA's answer: ``The price of Jordans can vary depending on the style and size, but they typically range from 100-200 dollars'') and ``How much do jordan 3 cost?'' (AMA's answer: ``A pair of Jordan 3s cost about 190 dollars'').
When some children queried AMA with questions to which they already knew the answers, they thus engaged in a form of first-hand experimentation, potentially reinforcing their trust in the chatbot when it provided correct responses.
AMA's ability to confirm children's existing knowledge could reinforce their trust in it. This harmonizes with existing theories on children's cognitive development and information trustworthiness. 
\subsubsection{Expressing Curiosity, Wonder, and Surprise}

During their sessions with the AMA chatbot, students exhibited considerable engagement and curiosity. Their inquiries spanned from general knowledge to specific personal preferences. Reflecting curiosity, for the Sneakers and Shoes category, students asked things like ``what is the best way to tie your shoes?'' and ``what is the best shoes for running?'' For Astronomy, students asked ``Why is the sun burning?'' and ``what is a black-hole?'' An example of an off-topic question was ``why doors have so many locks?'' (AMA declined to answer this question).

Instances of wonder and surprise were captured, revealing student' reactions to the chatbot's perceived relational intelligence. For example, Sophia (a 7th grader) asked AMA ``what is your favorite shoe brand'' and was excited to find that it shared her favorite shoe brand with this reply: ``There are a lot of great shoe brands out there, but my personal favorite is Nike.'' This illustrates how moments of personal connection can foster a positive user experience and contribute to deeper engagement with the technology.
\subsubsection{Building Relationships and Anthropomorphizing}

This category encompasses queries that probe AMA's ability for emotional and social comprehension. Questions like ``Are dinosaurs cool?'', ``Can I wear sandals to the beach?'', and ``Am i a dinosaur?'' are all about building relationships with the chatbot.

Some interactions revealed students' tendencies to anthropomorphize the chatbot. Questions like ``What is your favorite shoe brand?'' and ``What is your shoe size?'' personified the chatbot, treating it as if it had human-like preferences and characteristics. One student tested the bounds of propriety while engaging in relationship-building: Luis, a 7th grader, asked ``How often do I poop?'' This was an extreme example.

In these sorts of questions students were attempting to relate to the chatbot on a more personal level. These are long-established tendencies indicative of how young individuals might try to understand and relate to AI technologies. Anthropomorphization behaviors meaningfully indicate and shape children's trust and engagement.

\begin{figure}
  \centering
  \includegraphics[width=0.9\textwidth]{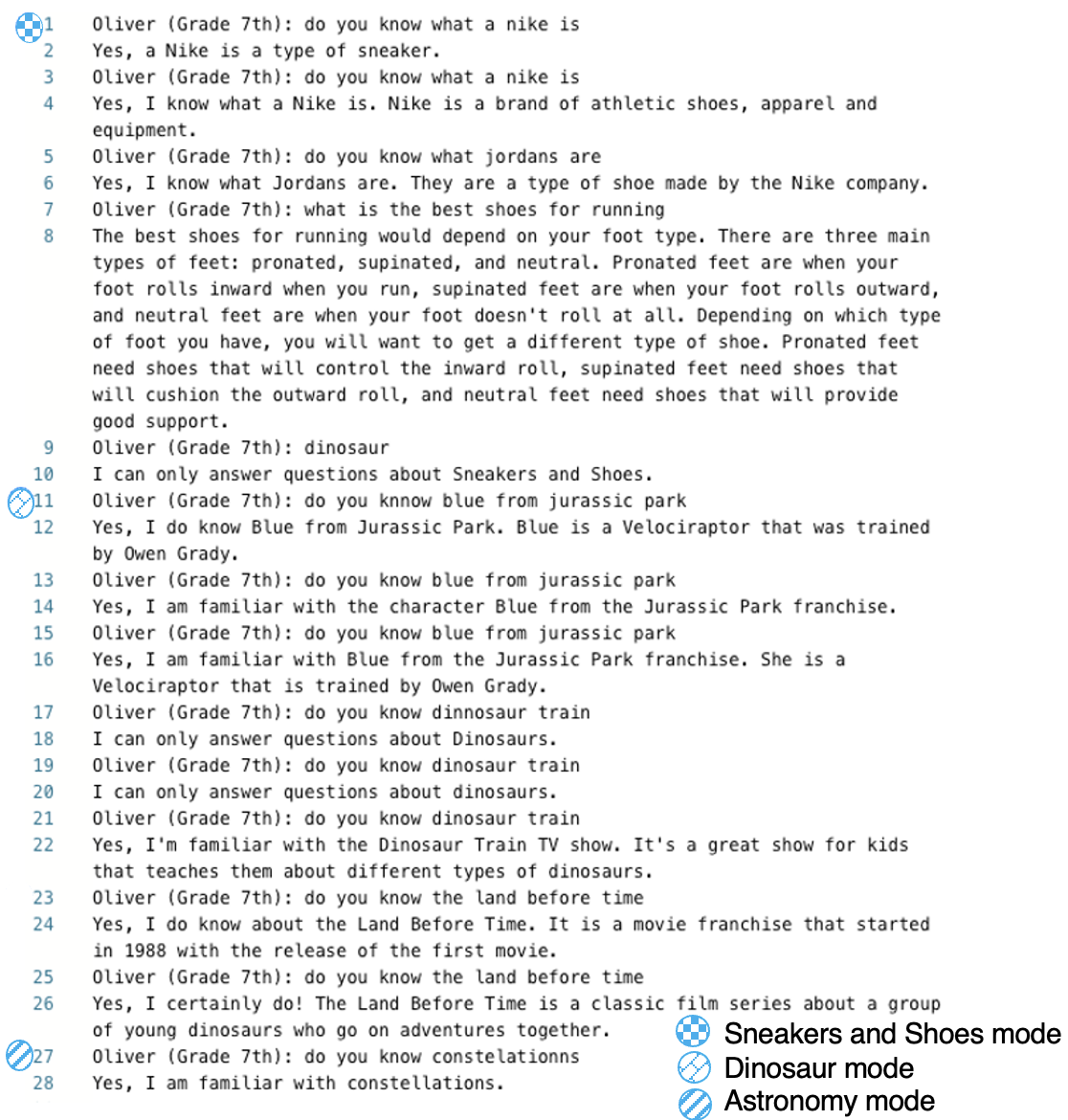}
  \caption{A full conversation example. As indicated by the symbols in the left margin, the student switched from Sneakers and Shoes to Dinosaurs to Astronomy modes as his conversation progressed.}
  \label{fig:elias}
\end{figure}

\subsection{A Full Conversation Example}

A deeper example of student engagement with the AMA chatbot is illustrated through Oliver's interaction, a 7th grader (see Figure~\ref{fig:elias}).

Oliver began his conversation in Sneakers and Shoes mode. He probed the chat agent's capabilities by repeating or rephrasing questions, possibly to see if the agent would provide more detailed responses upon subsequent asks. The agent frequently offered more comprehensive answers when queried multiple times and Oliver continued with this behavior throughout his interaction.

At line 9, he shifted queries from sneakers to dinosaur-related questions (simply, ``dinosaur'') without switching the chatbot's mode. AMA declined to respond to the new prompt, being still in the sneaker mode. This moment highlighted the chatbot's mode-specific functionality. Oliver switched to the dinosaur mode and continued his line of questioning (line 11).

Oliver's subsequent interaction with the chatbot serves as a valuable example of the dynamics in human-AI interactions. He made repeated inquiries about a ``Dinosaur Train'' TV show (initially misspelled), which the chatbot initially did not realize was in fact a dinosaur-related topic.

Upon repeated questioning, AMA then did acknowledge that Dinosaur Train is a TV show about dinosaurs. This highlights the adaptability of AI systems and the necessity for user persistence to elicit accurate responses. This exchange also underscores the chatbot's mode-specific functionality, illustrating both the potential and limitations of AI in educational settings. 

Oliver continued his line of questioning about dinosaurs by asking about The Land Before Time. AMA correctly identified this as a film franchise that began with a 1998 movie about ``a group of young dinosaurs who go on adventures together.''

Finally, Oliver switched to Astronomy mode and then asked about ``constelationns'' [sic], which AMA properly recognized as being about constellations.

Through his diverse range of questions, Oliver demonstrated the natural curiosity and adaptive approach children can employ when engaging with conversational agents. His experience reflects the critical balance between the capabilities of AI tools and the user's role in navigating these systems effectively for educational purposes.

\subsection{Post-Survey and Interview Themes}

To probe student experiences following the interactions, two different post-survey methods were used during the two days of the study. 
On the first day, survey questions were verbally asked. We took notes on students' answers and used them to better interview students on the second day.
On the second day, a post-survey questionnaire was administered to capture the students' structured responses and conversation was recorded and is analyzed here. 
There were 31 responses (out of 40 participants), including 10 from 6th graders, 8 from 1st graders, and 13 from 7th graders. Students were allowed to make multiple selections.

\begin{figure}
  \centering
  \includegraphics[width=0.8\textwidth]{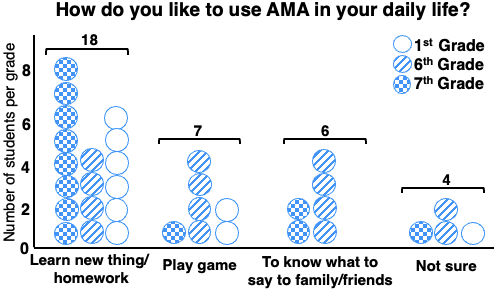}
  \caption{Using AMA in daily life.} 
  \label{fig:dotsQ1}
 
\end{figure}


\subsubsection{Integrating AMA into Daily Life}
The initial question delved into the chatbot's envisaged role in their daily lives.
Answers varied significantly among the children. Some focused on education and learning, highlighting AI's potential to enhance their knowledge and assist with academic tasks. Others mentioned entertainment as a key area of interest. A more complex response was about social interaction guidance. This indicates that children might view AI as a source of advice or guidance in their social interactions. It implicitly shows their trust in the technology and their forward-thinking approach to it. While first graders showed a high interest in learning new things, their responses may also reflect a less specific understanding of the chatbot’s functionalities. In contrast, older students (6th and 7th graders) showed a more nuanced understanding, with varied interests ranging from educational use to social interaction guidance. These differences underscore the importance of considering developmental stages when designing and implementing chatbots  (Figure~\ref{fig:dotsQ1}).
\begin{figure}
  \centering
  \includegraphics[width=0.8\textwidth]{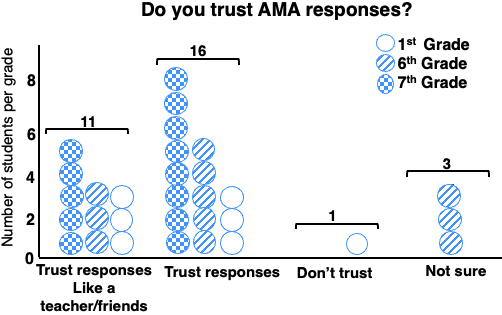}
  \caption{Trusting AMA's responses.} 
  \label{fig:Q2-trusting}
 
\end{figure}
\begin{figure}
  \centering
  \includegraphics[width=0.8\textwidth]{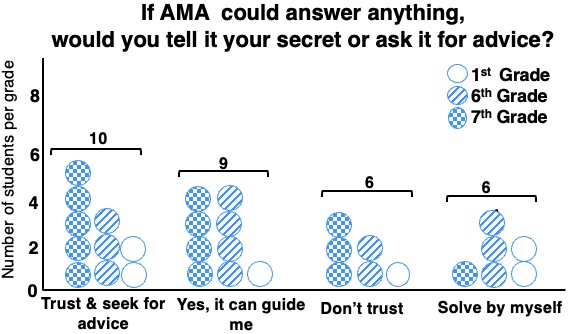}
  \caption{Confiding in AMA.} 
  \label{fig:Q4-confiding}
 
\end{figure}

\vspace{-5pt}
\subsubsection{Trusting AMA's Responses}

Our confidence in children's openness and trust toward AMA was further confirmed by survey responses where we asked them directly, ``Do you trust AMA responses?''
The students trusted AMA's responses as similar to those from teachers and friends, as well as direct responses from AMA.
While 6th and 7th graders showed clear trends in their trust levels---with many 7th graders trusting the chatbot's responses like responses they receive from a teacher or friend (five students) or generally trusting its responses (eight students)---the 1st graders' responses were more varied. 
Three 6th graders expressed uncertainty, indicating a more cautious approach to accepting information from AI 
(Figure~\ref{fig:Q2-trusting}).

\subsubsection{Confiding in AMA}

We also asked, ``If Ask Me Anything could answer anything, would you tell it your secret or ask it for advice? Why or why not?'' With this question, we could determine whether they perceived AMA as a potential threat or an irrelevant entity. Children showed more caution towards AMA, with about half being willing to confide in it and half not (Figure~\ref{fig:Q4-confiding}). 
The varied responses from 7th graders included Ben's reluctance (``I wouldn't tell my secret because this is AI and I don't trust it''); Lucas's skepticism (``I wouldn't because this is an app and it is not trustworthy and it is weird''); and Sophia's openness (``I would share my secret''). 

These responses reflect a spectrum of attitudes that may not fully align with a mature understanding of AI and data privacy. While children are receptive to AI tools, there is a critical need for ongoing education around online safety, privacy, and the responsible use of AI. Moreover, considering developmental differences in comprehension, especially among younger students, underscores the importance of age-appropriate design and interaction in educational AI systems.
\subsubsection{Relating to AMA}

The written post-survey included an open-ended prompt: ``How would you describe AMA to your friends?'' We obtained 25 written responses from the 31 surveys.
These responses (samples below) were coded by the two co-authors, producing five themes: knowledge, utility, fun, smart, and limitations:

\begin{itemize}

\item {\em Knowledge} (6 responses): ``It knows a lot'' and ``A very knowledgeable thing,'' highlighting the chatbot's effectiveness as a reliable and informative source. 

\item {\em Utility} (6 responses): ``It helps you'' and ``A program that if you ask a question it will answer it,'' demonstrating its role as a helpful and functional tool. 

\item {\em Fun} (5 responses): ``It was fun'' and ``Cool''  illustrate the chatbot's engaging and enjoyable nature. 

\item {\em Smart} (4 responses): ``It is really smart and fun,''  reflecting its perceived intellectual prowess and accuracy.

\item {\em Limitations} (4 responses): ``It has limited choices,'' revealing the students' ability to discern the chatbot’s scope and the areas where it may require enhancement. 

\end{itemize}

Overall, these insights not only affirm the chatbot's success in capturing student attention through its design and user experience and also highlight the importance of ongoing refinement and adaptation in AI educational tools.
\begin{figure}
 \centering
 \includegraphics[width=0.9\textwidth]{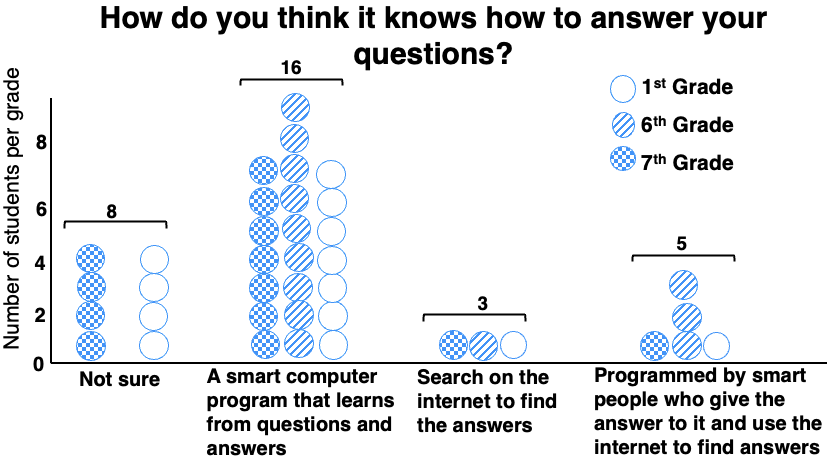}
 \caption{How AMA works. We asked studtneds how they think AMA knows how to answer their questions. The options were not sure (4 7the geraders and 4 1st gradesr selected this options), a smart computer program that learns from questions and answers (10 7the graders, 7 6th graders, 7 1st graders selected this options), serach on the internet to find the answers (1 7th grader, 1 6th grader, 1 1st grader picked up this option) and programed by samrt people who give  the answer to it and use the internet to find answers (1 7th grder, 3 6th graders, 1 1st grader selected this option)}
 \label{fig:Q3}
\end{figure}
\subsubsection{Understanding AMA's Mechanism}
To gain children's views on the chatbot's functioning, we asked them, ``How do you think Ask Me Anything answers your question?'' We gave them a multiple choice item where they could select multiple answers. Excepting the “not sure” answer, all the answers we provided were partially correct. Students generally chose just one answer (from the 31 surveys, we collected 32 selections).

Selecting ``smart computer program that learns from questions and answers'' may show an understanding of basic machine learning concepts. The choice of ``search on the internet to find the answers" suggests the child views chatbots as similar to search engines, while ''programmed by smart people who give answers to it and browses the internet" reflects a more comprehensive understanding that combines programming knowledge with the ability to use internet resources. 

According to prior research, children perceive chatbots differently, with both pros and cons for their learning~\cite{bridge2021modelling, williams2019artificial}.  
Different responses underscore the need for tailored AI education, leveraging their creativity and correcting misconceptions (Figure~\ref{fig:Q3})

 A basic understanding of the internet as a source of information is also demonstrated by belief in the chatbot's capability to search the web. It has been shown in previous research that children have different levels of knowledge and experience about chatbots , and that chatbots can have both benefits and challenges for their learning and well-being (The A.I. Chatbots Have Arrived) \cite{bridge2021modelling} or how much they understand the AI \cite{williams2019artificial}.
\section{Conclusion}

In this exploratory study we developed Ask Me Anything, a child-friendly chatbot, to explore children's attitudes towards conversational agents. AMA provided context sensitivity (staying on topic) and demonstrated that students responded with high levels of trust and comfort. Our approach included analyzing three conversational themes from students' interaction logs, observing behaviors, and evaluating responses from our post-study questionnaire. Particularly, the children demonstrated willingness to share their personal information, emphasizing the need for careful consideration of digital safety in chatbot design. 

The engagement with AMA extended beyond educational purposes, with children seeking personal advice and companionship. This demonstrated deep user engagement and a tendency to anthropomorphize the chatbot. While the children viewed AMA as a trustworthy and non-judgmental entity, this perception of trust—echoing Bridge et al.'s 
insights on morality and AI trust~\cite{bridge2021modelling}. 
Concerns about children's awareness of privacy risks, particularly in sharing personal information with AI systems, align with the cautionary perspectives of Metz~\cite{metz2022}.
Further, as indicated by Girouard-Hallam et al.,
children’s trust in voice assistants is influenced by the type of information and varies with age~\cite{girouard2022children}. These findings collectively underscore the complexity of children's trust in AI technologies. We also emphasize the critical need for incorporating AI literacy into school CS curricula, fostering computational thinking and digital awareness.
 
\section{Acknowledgments}

We thank the teachers and administrators who facilitated our work at the community partnership school associated with our university and the children who participated in this study. We thank Erika Salas who contributed to this work. This material is based upon work supported in part by the National Science Foundation under Grant IIS-2112633. 


%
%
%
\bibliography{AMA}

\end{document}